\documentclass[prd,reprint, nofootinbib,amsmath,amssymb, aps, floatfix]{revtex4-1}
\usepackage{dcolumn}

\usepackage{bm}
\pdfoutput=1
\usepackage{hyperref}
\usepackage{graphicx,subfigure}
\usepackage{epsfig}
\usepackage{amsfonts}
\usepackage[usenames]{color}
\usepackage{enumerate}
\usepackage[T1]{fontenc}

\usepackage{float}

\usepackage{xcolor}

\newcommand{\be}{\begin{equation}}
\newcommand{\ee}{\end{equation}}

\begin{document}
\title{Black hole entropy and the Bekenstein bound}
\author{Raphael Bousso}
 \email{Correspondence: bousso@lbl.gov}
\affiliation{Center for Theoretical Physics and Department of Physics\\
University of California, Berkeley, CA 94720, USA}
\affiliation{Lawrence Berkeley National Laboratory, Berkeley, CA 94720, USA}
\bibliographystyle{utphys-modified}

\begin{abstract}
I share some memories and offer a personal perspective on Jacob Bekenstein's legacy, focussing on black hole entropy and the Bekenstein bound. I summarize a number of fascinating recent developments that grew out of Bekenstein's pioneering contributions, from the Ryu-Takayanagi proposal to the Quantum Null Energy Condition.
\end{abstract}
\maketitle

\section{Personal memories}
\label{sec-memories}

Jacob Bekenstein entered my life as a formula; he left it as a cherished friend. 

I was Stephen Hawking's student, so I soon learned that black holes have entropy, according to Bekenstein's famous formula~\cite{Bek72,Bek73,Bek74}. Stephen and I studied the pair creation of black holes in cosmology. We used Euclidean methods to compute the (negligible) rate at which this process would happen. One expects configurations with less entropy to be more suppressed, so the Bekenstein-Hawking entropy of de Sitter and black hole horizons, though not central to our project, played a role as a consistency check. 

I was surprised to receive a postdoc offer from Bekenstein. We had not met personally, and none of my thesis work bore a close relation to his own interests at the time. My future plans leaned more towards cosmology than to the quantum aspects of black holes. My hope had been to spend time at Stanford and work with Andrei Linde. But I knew enough to understand that an offer from Bekenstein was something to be proud of. Also, I was relieved and grateful, because it was my only postdoc offer.

There was a problem: I was born in Israel, and though I had never lived there and do not speak Hebrew, I would have been subject to military service upon moving to Jerusalem. But Jacob was ready to help. After some bureaucratic wrangling, he obtained a permission for me to spend two years in Israel without being conscripted. Just then, with terrible timing, I learned that I had won a personal postdoctoral fellowship (by a German organization), with Stanford as the host institution. It was too late: I had already accepted Jacob's offer, and he had worked hard on my behalf to make it happen.

But when Jacob found out, he told me without hesitation that I should go to Stanford. I learned later that I was far from the only beneficiary of Jacob's habitual generosity and selflessness. He leaves behind a trail of gratitude.

Jacob's judgement was right. The physics questions he was grappling with at the time were profound, particularly the problem of a precise formulation of Bekenstein's nongravitational entropy bound, which I will discuss in Sec.~\ref{sec-bound}. But it would take more than a decade for a significant breakthrough to come about. At Stanford, I was able to make contributions on a shorter time-scale. I worked on cosmology with Linde, as I had hoped. I could easily visit Joe Polchinski in (relatively) nearby Santa Barbara. And from Lenny Susskind and a fascinating talk by Willy Fischler~\cite{FisSus98}, I learned about the problem of formulating a holographic entropy bound for cosmological spacetimes, which led to~\cite{CEB1,CEB2}. Ironically, Jacob had sent me straight into the beautiful web his work had spun, the fascinating connections between geometry and quantum information.

Though Jacob was personally kind and gentle, intellectually he was uncompromising and sometimes fierce. The question of whether nontrivial entropy bounds apply in general spacetimes was quite controversial in 1999. I was confident that I had resolved the issue, but it took some time for the debate to play out. Jacob was privately and publicly supportive, helping my work gain acceptance. His encouragement grew into a friendship and a source of strength over many years.

Jacob's colleagues did not allow his modesty to stand entirely in the way of proper recognition. It is a consolation to think back of the wonderful ``fest'' that his colleagues in Jerusalem insisted on organizing, allowing us to celebrate Jacob and his work just a few years before his untimely death. (He reluctantly agreed, but insisted that it be billed as a workshop.) Jacob received his share of prestigious prizes and awards. Yet, the full import of his most important insight, that black holes have entropy, is only beginning to dawn.

\section{Black hole entropy}
\label{sec-bhe}

Bekenstein's 1972 proposal that black holes carry entropy~\cite{Bek72} takes up four very small pages in a now extinct journal. It ranks as one of the most important physics papers of the 20th century. With 6 equations and less than 1500 words, it makes for an exceptionally quick and rewarding read.

Writing in concise, elegant prose, the author poses a puzzle: in the Second Law of thermodynamics, what accounts for the entropy of matter systems that disappeared into a black hole? In one sentence, he offers a proposal: black holes themselves must have some amount of entropy, such that the total (matter plus black hole) entropy according to an outside observer will never decrease.  In the next sentence, he makes the proposal quantitative: a black hole's entropy should be given by its horizon area in Planck units, up to an unknown factor of order unity. He then verifies that this proposal upholds the Second Law in a simple example. In four final lines he explains the signficance of what he has done: a counting of the quantum states of spacetime. Bekenstein's 1972 paper marks the first-ever quantitative statement intrinsically about quantum gravity. 

The most remarkable aspect of this revolutionary work (aside from the fact that it was written by a graduate student) is the premise from which it sets out. Bekenstein argues that we should {\em worry} about the loss of matter entropy into a black hole, because it would make it impossible to {\em verify} that the Second Law of thermodynamics still holds.\footnote{This problem was first posed to Bekenstein by his advisor, John Wheeler~\cite{Bek80}. For a fascinating first-hand account of these events, see Bob Wald's contribution to this memorial volume~\cite{Wal18}.}

A lesser mind might have been content with making excuses along the following lines. {\em Well, bad things happen all the time. I'm sorry that your box of gas fell into a black hole. It may be a bit awkward to visit your box there and to see how its entropy is doing. But this is an inconvenience, not a fundamental problem that needs solving. Probably the entropy ends up at the singularity inside the black hole, and we do not have a theory yet that describes the singularity. But if we had a full-fledged quantum gravity theory, we could account for the entropy in this way.} Bekenstein, by contrast, stood firm in insisting that the laws of physics continue to make sense in our laboratories, whether or not a black hole forms in one of the test tubes.

An observer who chooses to remain outside cannot access matter inside the black hole. And by the definition of ``black hole,'' she cannot see the matter or receive signals from it while she remains outside. General relativity further dictates that by jumping into a black hole, she cannot ``catch up'' with the matter that formed it. Indeed, she cannot even receive a signal from any matter that entered the black hole more than a time $(R/c)\log (R/L_P)$ before she does. Here times are measured at some distant radius from which the infall starts; $R$ is the radius of the black hole and $L_P=(G\hbar/c^3)^{1/2}$ is the Planck length, about $10^{-33}$ cm.\footnote{In the remainder of this article, I will set $c=1$ but keep Newton's constant $G$ and Planck's constant $\hbar$ explicit.}

It was a major leap for Bekenstein to elevate these obstructions to a crisis for the Second Law of thermodynamics. In insisting that the Second Law should retain operational meaning, Bekenstein was perhaps guided by the fact that black holes do carry gauge charges such as energy or electric charge. Their conservation can be verified at all times from the outside. He was familiar, moreover, with Hawking's recent proof of a theorem in general relativity: with reasonable (though ultimately invalid) assumptions on the positivity of the energy of matter, the area of all event horizons cannot decrease. There is a superficial analogy between this area theorem and the Second Law, which states that the total entropy cannot decrease.

But the Second Law is just a statistical probability, whereas the area theorem, within classical general relativity, was a mathematical certainty. This did not deter Bekenstein: his proposal was that the analogy is an identity. He thus asserted that the area of a geometric surface is more fundamentally a statistical quantity. Nearly half a century later, this remains the single deepest insight we have gained into the fabric of space and time. It has proven extraordinarily fruitful and lies at the core of today's most promising avenues for studying quantum gravity. 

Whereas Bekenstein advocated that the area of a black hole should be treated as a real contribution to the total entropy, it is interesting that he did not go further and ascribe a temperature to black holes. In fact, his proposal implied (by the First Law of thermodynamics) that black holes have a temperature and hence must radiate. Rather ironically, this implication prompted Hawking to reject the proposal at first. Classically, nothing can come out of a black hole, so it seemed absurd to think that black holes could radiate. But of course, Bekenstein's proposal is quantum mechanical in nature, so a purely classical argument against its implications carries no weight. In 1974 Hawking discovered black hole radiation as a quantum effect, in an independent calculation. Working back via the First Law, this fixed the proportionality factor in Bekenstein's proposal to be $1/4$.

\section{Generalized Entropy and Generalized Second Law}
\label{sec-gsl}

The terms ``generalized entropy'' and ``generalized second law'' first appear in Bekenstein's 1973 paper~\cite{Bek73}. We now think of them as fixed names for important concepts, but in this first instance, the word ``generalized'' is merely descriptive, a convenient way of referencing the notions introduced in the 1972 paper. Bekenstein distinguishes the ``common'' entropy of the matter systems outside the black hole from the ``generalized'' entropy. The latter includes both the black hole contribution and the matter entropy:
\begin{equation}
  S_{\rm gen}= \frac{A}{4G\hbar} + S_{\rm out}~.
\label{eq-sg}
\end{equation}

The generalized second law (GSL) is Bekenstein's conjecture that the generalized entropy satisfies a Second Law of thermodynamics:
\begin{equation}
  d S_{\rm gen} \geq 0~.
\end{equation}
It took nearly four decades for the GSL to be proven in a wide regime. Using methods from algebraic quantum field theory, and exploiting specific properties of null quantization, Wall~\cite{Wal11} proved the GSL perturbatively in $G\hbar$, for free fields. The proof extended automatically to interacting fields, after it was established that their modular Hamiltonian takes the same form on such null hypersurfaces~\cite{KoeLei17,CasTes17}. Beyond the perturbative expansion in $G\hbar$, the GSL remains a conjecture.

Thanks to the success of Bekenstein's proposal, the modifier ``generalized'' is now outdated.\footnote{I thank Stefan Leichenauer for frequently complaining about this.} The theoretical evidence that black holes have entropy is overwhelming. It would be silly to exclude this contribution from the total entropy and expect the Second Law to remain valid. Thus, generalized entropy is simply the entropy. 

But there are reasons to keep the name. The sum $S_\text{gen}$ is actually better defined than either of its two separate terms on the right hand side of Eq.~(\ref{eq-sg}). In quantum field theory, $S_\text{out}$ can be defined as the von Neumann entropy of the quantum state restricted to the exterior of the horizon. This has divergences; the leading divergence is proportional to the area in units of the UV cutoff. Newton's constant $G$ also has a nontrivial renormalization group flow. There is strong evidence that these flows cancel, so that the sum $S_\text{gen}$ is finite and well-defined. (See the appendix of Ref.~\cite{BouFis15a} for a detailed discussion and extensive references.) There are subleading divergences that cancel as well~\cite{Wal93,IyeWal95,JacKan94,Don13}.

Moreover, there are a number of theorems in general relativity which hold only if one assumes the null energy condition (NEC) on the stress tensor, that $T_{ab} k^a k^b\geq 0$ for any null vector $k^a$. Hawking's area theorem is an example. But in fact, the null energy condition can be violated by quantum fields. An important example of this is Hawking radiation: the black hole shrinks, in violation of the area theorem. 

The generalized entropy can be viewed as a quantum-corrected area, since the area term dominates by one power of $\hbar$ over the matter entropy term.  And unlike the area, the generalized entropy {\em does} increase during black hole evaporation: it can be shown that the entropy of the Hawking radiation more than compensates for the lost area~\cite{Pag76}. Indeed, this behavior is required by the GSL.

We see that there are really two ways of looking at Bekenstein's generalized entropy. The first is the way he discovered it, as a black hole's contribution to the total entropy, and thus as a way to save thermodynamics. But we can also think of generalized entropy as a kind of quantum-corrected area that ``saves'' a theorem of general relativity. Hawking's area theorem fails in the presence of quantum effects, but the analogous statement for generalized entropy survives.

Moreover, $S_\text{gen}$ can be associated not only to slices of a black hole event horizon, but to arbitrary spatial surfaces that split a Cauchy surface into two parts. For example, we can define a generalized entropy of the Earth's surface, given by its area plus the von Neumann entropy of its exterior (or interior). This broader notion of generalized entropy was not envisaged in Bekenstein's original proposal. But in recent years, it has proven exceptionally powerful.

Consider any GR theorem that involves the area of surfaces or some derivative thereof. Suppose, moreover, that the theorem relies on the NEC and hence breaks down in NEC-violating quantum states. Since we can view the generalized entropy as a quantum-corrected area, it is natural to ask whether we can ``save'' the GR theorem in the presence of quantum effects, by replacing area with generalized entropy in the statement of the theorem. This trick works for Hawking's area theorem, so why not apply it in other contexts?

For example, Penrose's theorem~\cite{Pen64} guarantees a singularity in the presence of a trapped surface, assuming the NEC. But the NEC and hence Penrose's theorem can fail. It is thus not clear whether the inevitability of singularities and the associated breakdown of general relativity is merely an artifact of a classical limit. But Wall~\cite{Wal10} used Eq.~(\ref{eq-sg}) to define the notion of a ``quantum-trapped'' surface. He further showed that the GSL implies a singularity in the future of any such surface. Thus, Penrose's singularity theorem can be transferred to firmer ground, onto the foundation laid by Bekenstein in 1972.

Another example is the area theorem for holographic screens~\cite{BouEng15a,BouEng15b}, which applies to surfaces in cosmology and inside black holes. By judiciously replacing area with generalized entropy in the analysis, this result can be elevated to a GSL for ``Q-screens''~\cite{BouEng15c}, and thus to a well-defined statement of the Second Law in cosmology.

Even the classical focussing theorem, which governs the evolution of cross-sectional area spanned by light-rays, can be elevated to a ``quantum focussing conjecture,'' again by appropriately replacing area with generalized entropy. The QFC turns out to be a very powerful statement, to which we will return later.



\section{Bekenstein bound}
\label{sec-bound}

\subsection{From the GSL to $S\leq 2\pi ER/\hbar$}
\label{sec-bbound}

If black holes did not carry entropy, then as Bekenstein noted, it is inevitable that the Second Law of thermodynamics can be violated from the point of view of an external observer, because matter entropy can be lost into the black hole. 

However, even if black holes do carry entropy as Bekenstein proposed, $S_{BH}=A/4G\hbar$, it is not immediately clear that the Second Law is saved. This would appear to depend on how the black hole responds to the addition of matter systems that carry entropy. The question is whether the area of a black hole increases {\em by enough} to make up for the lost entropy, or not.

The growth of the horizon area is controlled by the Einstein equation of general relativity, which involves geometric quantities and the stress-energy tensor of matter. Thus the area increase will manifestly depend on the size and energy density of the matter system, and on its path across the horizon.

From this viewpoint, it is completely unclear why the horizon area increase would be sensitive to the entropy of the matter system, i.e., to the number of its possible quantum states. But the GSL demands this. Thus the GSL carries within it a fascinating, deep connection between quantum information on the one hand, and energy and geometry on the other.

A stationary black hole is entirely characterized by its conserved charges: its electric charge, its angular momentum, and most importantly (since it limits the previous two), its mass. The area of its event horizon, therefore, is a function of only these three quantities. It has no manifest dependence on the entropy of the matter that entered the black hole (that is, on the log of the number of quantum states that this matter could have been in). So how can the area possibly ``know'' that it should increase enough to compensate for this lost matter entropy?

To make this concrete, suppose that we add a neutral matter system with no angular momentum to an unspinning neutral black hole. Thus, the black hole is described by the Schwarzschild solution with mass $M$ before and mass $M+\delta M$ after the addition of the system. We can take the black hole to be large, so that $M\gg \delta M$. The Schwarzschild radius is proportional to the mass, so the area of the horizon increases in this process, by $\delta A = 32 \pi G^2 M\delta M$. Hence, the entropy of the black hole will increase by $\delta S_{BH} = 8\pi G M\delta M/\hbar$.

In order to violate the GSL in this example, we would have to arrange for $\delta S_{BH}$ to be smaller than the entropy $S$ of the matter system we are adding to the black hole. This means we have to keep $\delta M$ small. That is, we would like the matter system to add as little energy as possible to the black hole. This is best accomplished by extracting work from the entropy-carrying matter system, by lowering it slowly towards the horizon of the black hole (using a string, say), and only then dumping it in. By this trick, we effectively lower the mass of the system by redshifting it into a gravitational well. 

It might seem that we can lower the system all the way to the horizon, where the redshift diverges. If this was possible, then we could extract all of its rest mass as work. Then the black hole mass would not increase at all once the matter system is finally dropped in: $\delta M=0$. We could violate the GSL: the generalized entropy would decrease, because the black hole area would remain fixed same whereas the matter entropy would have disappeared.

But Bekenstein realized~\cite{Bek74,Bek81} that this is impossible: any matter system will have finite size, so we cannot lower its center of mass arbitrarily close to the horizon without tidal disruption of the system. For concreteness, consider a spherical box with radius $R$ and roughly homogeneous mass distribution. We must be content to lower its center of mass to a proper distance $R$ from the horizon. There is no more work we can extract, without having parts of the system already enter the black hole.

A bit of algebra shows that in this optimal case, $\delta M = ER/4GM$. (See, e.g., Ref.~\cite{RMP} for a brief derivation.) Thus, the black hole area will increase by $\delta A = 8\pi GER$, and the black hole entropy will increase by $\delta S_{BH} = 2\pi ER/\hbar$. The GSL requires that $\delta S_{\rm gen}\geq 0$ and hence that $\delta S_{BH}>-\delta S_{\rm out}=S$. By insisting on the validity of the GSL, we are thus led to the Bekenstein bound:
\begin{equation}
  S\leq 2\pi ER/\hbar~,
  \label{eq-bekbound}
\end{equation}
where $S$, $E$, and $R$ are the entropy, energy, and radius of the matter system.

It is very significant that Newton's constant $G$ does not appear in this final result. Neither do any parameters that describe the black hole. The Bekenstein bound is purely a nongravitational statement, and it is intrinsically about the matter system itself: the first universal entropy bound.

\subsection{Controversy}
\label{sec-controversy}

The Bekenstein bound, Eq.~(\ref{eq-bekbound}), was at the center of a heated controversy in the 1980s and 1990s. Ref.~\cite{Bek99} has many references and a nice summary of the back and forth, which traces back at least to~\cite{UnrWal83}. (To my knowledge, the GSL itself was not controversial.) The debate centered largely on two questions: (i) Is Eq.~(\ref{eq-bekbound}) indeed necessary for the GSL? That is, does the Bekenstein bound follow rigorously from the GSL, rather than just being inspired by the handwaving arguments of the previous subsection?  (ii) Is Eq.~(\ref{eq-bekbound}) true? Or is it possible to find matter systems whose entropy exceeds the Bekenstein bound?

Both were valid questions. The thought-experiment of lowering the system slowly towards the horizon, which links the GSL to the Bekenstein bound, is beset by subtleties and potential loopholes. Thus, it appeared logically possible for the bound to fail even if the GSL holds, or for the bound to be true but not to be implied by the GSL.

Concerning (i), Unruh and Wald~\cite{UnrWal83} led the charge in arguing that known properties of quantum field theory were sufficient to protect the GSL. Thus the GSL would not impose new constraints such as Eq.~(\ref{eq-bekbound}). (See also Ref.~\cite{Wal18}.) I will comment on this viewpoint further at the end of Sec.~\ref{sec-qnec}.

Regarding (ii), there seemed to be fairly obvious counterexamples to the Bekenstein bound. For example, certain quantum fields have negative Casimir energy in certain cavities. Thus $E<0$ and $S=0$ in the ground state of the system. 

In hindsight, the question we should have been focussing on was: (iii) Exactly how should one define $S$, $E$, and $R$ in Eq.~(\ref{eq-bekbound})? In fact, this question did receive some attention, but until 2008, there were no satisfactory proposals.

Bekenstein argued that $E$ must include all necessary parts of the system, by which he meant a ``complete system'' that ``could be dropped whole into a black hole''~\cite{Bek00b}. This gets rid of the Casimir counterexample, because the rest mass of the walls of the cavity overcompensates for the negative energy of the ground state. However, in forcing us to include the rest energy of enclosures, Bekenstein risked trivializing his bound, by making it virtually impossible to find a system that would come close to saturating it.

Bekenstein also insisted that for nonspherical systems, $R$ should be defined as the radius of the smallest circumscribing sphere. This helped fend off potential counterexamples, but it weakened the bound for no principled reason. In the thought-experiment that led to the bound, in Sec.~\ref{sec-bbound}, we are free to orient a thin slab so that its long side is parallel to the horizon. Thus, the shortest dimension determines how close we can get it to the horizon before dropping it in. This yields a bound of the form
\begin{equation}
  S\leq \pi Ew/\hbar~,
  \label{eq-tight}
\end{equation}
where $w$ is the smallest separation of two infinite parellel plates between which we can fit the system. As pointed out to me by Casini, the ``necessary parts'' argument alone would suffice to eliminate the apparent counterexamples to this stronger bound, so it was not clear why Bekenstein went with the weaker version.\footnote{In the conversations we had about this, Jacob was characteristically low-key, offering only that he intended the bound as a ``rule of thumb.'' It was an interesting contrast to the intensity with which he pursued these debates in scientific publications.}

\subsection{Relation to the covariant entropy bound}
\label{sec-relation}

In 2002, I was able to show~\cite{Bou03} that the covariant entropy bound~\cite{CEB1}, in its generalized form~\cite{FMW}, implied the ``tight'' version of Bekenstein's bound, Eq.~(\ref{eq-tight}), in a weakly gravitating setting. The generalized covariant entropy bound states that
\begin{equation}
  S_\text{matter}\leq \frac{A-A'}{4G\hbar}~.
\label{eq-gceb}
\end{equation}
Here $S_\text{matter}$ is the entropy of the matter systems crossing a light-sheet of an arbitrary spatial surface $\Sigma$. A light-sheet is a family of non-expanding null geodesics (light-rays) that emanate orthogonally from $\Sigma$.  $A$ and $A'$ are the areas of $\Sigma$ and of the final surface on which the light-sheet is terminated, respectively. 

The general argument that Eq.~(\ref{eq-gceb}) implies Eq.~(\ref{eq-tight}) is quite simple. For this brief summary, let us make it even simpler by restricting to an approximately spherical matter system of radius $R$, and neglecting factors of order unity.

Consider a congruence of parallel light-rays in Minkowski space, such as would emanate from lasers mounted on a flat wall at $x=0$ that all flash for an instant at $t=0$. These light-rays occupy the null plane $x=t$. (Here, light-rays does not mean actual light, which might be blocked or deflected. We consider null geodesics, which pass through any matter system like X-rays and which are only affected by gravity.)

The matter system presents a cross-section of area $\sim R^2$ to the light-rays. In passing through the matter system, the geodesics will be focussed by an angle $\phi\sim GE/R$. (This is the same gravitational effect that is responsible for the famous bending of light by the sun.) This angle will be small by the assumption that gravity is weak, i.e., because our matter system is something like planet Earth, not a neutron star or an expanding universe. To first order in $G$, therefore, the cross-section spanned by the light-rays will have decreased by an annulus of width $\sim \phi R$ and hence of area
\begin{equation}
  \Delta A \sim \phi R^2 \sim GER~.
\end{equation}
The generalized covariant entropy bound states that the entropy of the matter system satisfies $S\leq \Delta A/4G\hbar$, which yields the Bekenstein bound. A more careful analysis produces the tighter version of the bound, Eq.~(\ref{eq-tight}), including the prefactor $\pi$.

This was a pleasing result. It meant that Bekenstein's original 1981 bound did not stand logically apart from the ``holographic'' entropy bound I had found much later. Instead, it corresponded to a particular weakly gravitating limit of a more general, gravitational bound.

But this result also meant that the generalized covariant entropy bound was subject to the same questions as Bekenstein's bound. What about apparent violations such as Casimir energy? In some cases $\Delta A$ is better defined than $E$ and $R$ separately, but it remained unclear what exactly constitutes an ``isolated'' and ``complete'' matter system. Where does it begin and end?

We now know that this was the wrong question. One should simply pick a (null) region that will be considered, and compute its generalized entropy~\cite{BouFis15a}, or in the nongravitational limit, the vacuum-subtracted von Neumann entropy of the quantum field theory state restricted to this region~\cite{Cas08,BouCas14a}. But it did not occur to me to use the generalized entropy for non-horizon surfaces, and I was deterred by the fact that the von Neumann entropy is divergent, due to the entanglement entropy across the boundary of the region. So, in a set of deservedly forgotten papers, I pursued various other approaches to sharpening the formulation of the Bekenstein bound~\cite{Bou03a,Bou03b}.

My goal was to answer question (iii), and to formulate the Bekenstein bound in a way that was neither wrong, nor trivial, nor ill-defined. The key challenge I focussed on was to define the size of a system, given that quantum wave functions are usually not sharply localized. I felt I was making progress, but I did not reach a fully satisfactory solution, and in fact I was on the wrong track.

But the problem did have a solution. Its discovery by Casini would form the seed of several recent breakthroughs that are significantly deepening our understanding of the connection between quantum information, energy, and spacetime.

\subsection{Marolf, Minic, and Ross}
\label{sec-mmr}

In 2003 Marolf, Minic, and Ross~\cite{MarMin04} (MMR) pointed out a curious property of the entropy one can ascribe to a matter system. As they put it, it is ``observer-dependent.'' Their insight built on earlier work by Marolf, Sorkin, Unruh, Wald, and others, as cited in~\cite{MarMin04}.

Suppose that a field excitation with fixed energy $E$ and spatial size $R$ can be in a large number $n$ of different quantum states. This would be possible, for example, in a theory with $n$ different species of photon. We can then prepare a maximally mixed state of all possible species. That is, we know that a photon wavepacket is present, but not which species it was made from. The entropy of this wavepacket is $\log n$. If $n$ is large enough, so that $\log n > 2\pi ER$, then the entropy will exceed the Bekenstein bound.

This ``species problem''  had long been recognized as presenting a challenge. A common response was to note that the actual number of light species in Nature  cannot be freely chosen, and that it happens to be quite small. One might reason that  this fact is no accident, but the result of the Bekenstein bound constraining the number of species.

But MMR noticed that the entropy of such a wave packet would not in fact grow like $\log n$, from the point of view of an accelerated observer who had access only to one side of a Rindler horizon. This was significant, since an observer hovering near the horizon of a black hole is locally like a Rindler observer. 

MMR computed the entropy only in a certain limit of small excitations above the vacuum. But the calculation was done in full quantum field theory: they considered the actual density operators that describe the vacuum and the excited state, restricted to the Rindler patch ($x>0$ at $t=0$). The von Neumann entropy of either state diverges, due to entanglement across the boundary at $x=0$. As usual, one would like to say that this divergent part is not important, and that instead that one should focus on the entropy of the excitation. But what formula should one attach to these words?

MMR reasoned that one could simply compare the divergent entropies with and without the excitation. The difference would capture the entropy that can be attributed to the excitation. Indeed, for ordinary matter systems localized well inside the Rindler region, this quantum field theory prescription reproduces the standard thermodynamic entropy (say, of a box of gas).

Written as a density operator, the global vacuum is the state $|0\rangle\langle 0|$. Restricting the vacuum state to the right Rindler patch ($x>0$) means tracing out the left patch:
\begin{equation}
  \sigma=\text{Tr}_L\, |0\rangle\langle 0|~.
\end{equation}
Similarly, one can chose a global excited state $\rho_g$ and define its restriction to the Rindler patch as
\begin{equation}
  \rho = \text{Tr}_L\, \rho_g~.
\label{eq-rhorhog}
\end{equation}
The (divergent) von Neumann entropies of these states are $-\text{Tr}\sigma\log\sigma$ and $-\text{Tr}\rho\log\rho$, respectively. MMR's analysis was restricted to states such that $\rho=\sigma+\delta \rho$, with $\delta \rho\ll 1$. Working to first order in $\delta\rho$, the difference between the entropies is 
\begin{equation}
  \Delta S = -\text{Tr\,} [\delta \rho\, (1+\log \sigma)] = -\text{Tr}[\delta\rho\, \log\sigma]~.
  \label{eq-edr}
\end{equation}

Crucially, MMR found that the entropy difference in the Rindler patch, Eq.~(\ref{eq-edr}), does not grow as $\log n$ indefinitely, as it would if we computed it from the global quantum state. In the limit of a large number $n$ of internal states, at fixed global energy and size of the system, the Rindler-patch entropy difference {\em saturates} at
\begin{equation}
  \Delta S = 2\pi \,\delta E_R/\hbar~,
\label{eq-mmr}
\end{equation}
where
\begin{equation}
  \delta E_R=\int dx\, dy\, dz\, x\, T_{tt}~,
\end{equation}
is the Rindler energy and $T_{tt}$ is the energy density. 

In the special case where the system is sharply localized compared to its distance $R$ from the Rindler horizon, the Rindler energy is simply $\delta E_R \approx E R$, where $E = \int dx\, dy\, dz\, T_{tt}$ is the global energy. The result of MMR, Eq.~(\ref{eq-mmr}), thus implies that in the limit of diverging species number $n$, the Bekenstein bound becomes saturated but not violated. This holds at least in the limit studied by MMR, if the entropy is defined as they proposed, from the viewpoint of the Rindler observer. 

But the Rindler horizon could secretly be a black hole horizon, just as the Earth's surface looks flat to us. Expressed in terms of asymptotic energy and black hole temperature, the MMR result is that in the large $n$ limit, the entropy of an excitation just outside a black hole saturates at $\delta E/T_{BH}$. By the First Law, this is precisely the amount by which the black hole entropy will increase if we drop in the system. Thus, MMR found that the GSL would be satisfied, and saturated, in this limit.

\subsection{Casini's formulation and proof of the Bekenstein bound}
\label{sec-casini}

For the results of MMR, described above, it is crucial that the entropy be defined as a vacuum-subtracted entropy. This is the difference between the von Neumann entropy of the excited state restricted to the Rindler region, $\rho$ (or to the outside of the black hole), and that of the vacuum state so restricted, $\sigma$. In 2008, Casini~\cite{Cas08} considered the vacuum-subtracted entropy of arbitrary states in the Rindler region. (Thus he went beyond a first-order analysis of states that were close to the vacuum.) He used tools from algebraic quantum field theory to state and prove a rigorously defined version of Bekenstein's bound.

Casini pointed out that
\begin{equation}
  S_{\rm rel} (\rho|\sigma) = \Delta K - \Delta S~,
  \label{eq-srelsplit}
\end{equation}
which follows immediately from the definitions of the relative entropy,
\begin{equation}
S_{\rm rel} = \text{Tr}\, \rho\log\rho-   \text{Tr}\, \rho\log\sigma ~,
\end{equation}
the vacuum-subtracted entropy
\begin{equation}
\Delta S = -\text{Tr}\, \rho\log\rho +\text{Tr}\, \sigma\log \sigma~,
\end{equation}
and the modular energy $\Delta K$,
\begin{equation}
\Delta K = \langle \rho | K_\sigma|\rho \rangle - \langle \sigma | K_\sigma| \sigma\rangle~.
\end{equation}
Here the modular Hamiltonian $K_\sigma$ of the vacuum state $\sigma$ is defined via the operator equation
\begin{equation}
\sigma = \frac{e^{-K_\sigma}}{\text{Tr}\, e^{-K_\sigma}}~.
\end{equation}

In general, the modular Hamiltonian is not an integral of a local density. But in the case at hand, $\sigma$ is the vacuum restricted to the Rindler patch. This implies that $\Delta K$ is proportional to the Rindler energy~\cite{BisWic76}:
\begin{equation}
  \Delta K = \frac{2\pi}{\hbar} \int dx\,dy\,dz\,x\,T_{tt}~.
\end{equation}
As noted earlier, for objects that are well-localized at a distance $R$ from the Rindler horizon, this implies that $\Delta K\approx 2\pi ER/\hbar$.

These observations motivated Casini to reformulate the Bekenstein bound as the claim that
\begin{equation}
  \Delta S \leq \Delta K~.
  \label{eq-casinibound}
\end{equation}
This statement is well defined for all finite-energy quantum states, not just those for which we can give an intuitive definition of $E$ and $R$. For the latter, it reduces to $\Delta S\leq 2\pi ER/\hbar$. Thus Casini solved the decades-old problem of giving a sharp formulation of the Bekenstein bound.

Casini then proved this bound, in one line: Eq.~(\ref{eq-casinibound}) follows from Eq.~(\ref{eq-srelsplit}) and the known property that the relative entropy is nonnegative, $S_\text{rel} \geq 0$.

\section{Recent developments}
\label{sec-recent}

Bekenstein's notion of generalized entropy lies at the heart of several important recent developments. All of them are explore the concept beyond its original setting, applying it to any surface $\Sigma$ that splits the world\footnote{By ``the world,'' I mean a Cauchy surface.} into two sides. It does not matter whether $\Sigma$ lies on a black hole event horizon. As noted in Sec.~\ref{sec-gsl}, in this context it can be useful to think of generalized entropy as the ``quantum-corrected area'' of the surface $\Sigma$:
\begin{equation}
  A_Q = A+ 4G\hbar\, S_\text{out}~,
\end{equation}
where
\begin{equation}
  S_\text{out} = - \text{Tr}\, \rho \log \rho
\end{equation}
is the von Neumann entropy of the state $\rho$ of the quantum fields restricted to one side of the surface, and $\rho$ is given by a partial trace of the global state over the other side as in Eq.~(\ref{eq-rhorhog}). More refined algebraic definitions of $S_\text{out}$ deal with the divergent entanglement structure~\cite{Haa92,Wit18} and with subtleties arising from gauge constraints~\cite{CasHue13,DonWal14,DonWal15}.

\subsection{Ryu-Takayanagi proposal}

In the context of the AdS/CFT correspondence~\cite{Mal97}, Ryu and Takayanagi proposed in 2006 that the von Neumann entropy of any boundary subregion is given by the area of the minimal bulk surface $\Sigma$ homologous to this region~\cite{RyuTak06}:
\begin{equation}
S_\text{bdy} = \frac{A[\Sigma]}{4G\hbar}
\end{equation}
Hubeny, Rangamani and Takayanagi soon extended the proposal from static to general settings, where there exist no minimal surfaces. Instead, the area of the smallest stationary\footnote{By ``stationary,'' I mean a surface whose area does not change to first order in small deformations. In the AdS/CFT literature this is more frequently referred to as an ``extremal'' surface. But ``extremal'' means minimal or maximal. In a Lorentzian geometry, stationary spatial surfaces can only be saddle points of the area functional.} homologous bulk surface~\cite{HubRan07} computes the entropy. A useful ``maximin'' reformulation of this proposal was given by Wall~\cite{Wal12}.

Faulkner, Lewkowycz, and Maldacena~\cite{FauLew13} developed the proposal further, going beyond leading order in $1/N$ (or equivalently, in $G$), where $N$ is the rank of the gauge group of the boundary field theory. They argued that the boundary entropy is more accurately computed up to $O(N^0)$ by adding the ordinary matter entropy in the bulk to the area. Interestingly, this was not formulated explicitly as the generalized entropy in the sense of Bekenstein (Sec.~\ref{sec-gsl}).

This connection was made clear, and the proposal extended to all orders in $1/N$, by Engelhardt and Wall~\cite{EngWal14}. They suggested that the relevant surface $\Sigma$ should be defined as the ``quantum extremal surface'' (really, quantum stationary surface), for which $A_Q$ is locally stationary. The bulk generalized entropy $A_Q$ then computes the boundary entropy:
\begin{equation}
S_\text{bdy} = \frac{A_Q[\Sigma]}{4G\hbar}~.
\end{equation}

These investigations helped resolve the problem of subregion-subregion duality~\cite{BouLei12,BouFre12}. Refs.~\cite{Wal12,HubRan12,CzeKar12} argued that the dual to any boundary region is the entanglement wedge, i.e., the region enclosed by the boundary region and the bulk surface that computes the entropy. This is now well established~\cite{JafLew15,DonHar16}. It has led to beautiful insights concerning the role of quantum error correction in the emergence of the bulk spacetime~\cite{AlmDon14,Har16}.

\subsection{Cosmological area theorem and Second Law}

The notion of a holographic screen was first made precise in~\cite{CEB2}. Given an arbitrary spacetime, a holographic screen can be non-uniquely constructed by choosing a null foliation of the spacetime. The screen is the hypersurface consisting of a one-parameter family of marginally trapped or anti-trapped surfaces called leaves. (In some cases, like asymptotically flat or AdS spacetime, the screen can reside on the conformal boundary.)

This definition was chosen so that the covariant entropy bound~\cite{CEB1}, applied to each leaf, constrains the entropy on the entire null slice that it is embedded in, in terms of the area of the leaf. In other words, holographic screens make manifest the idea~\cite{Tho93,Sus95} that the world is a hologram. In asymptotically AdS spacetimes, we have an actual quantum gravity theory supporting this conjecture. But the covariant entropy bound applies to any spacetime, including the universe we live in, suggesting that the holographic nature of quantum gravity is universal. For a review see~\cite{RMP}.

In Ref.~\cite{BouEng15a,BouEng15b}, Engelhardt and I distinguished between past and future holographic screens, which consist respectively of marginally trapped and marginally antitrapped surfaces. (These objects are similar to but more general than ``dynamical horizons''~\cite{AshKri04}.) We further proved that any past or future holographic screen satisfies a nontrivial area theorem: the area of the leaves increases monotonically along the screen. This is distinct from Hawking's area theorem, which applies to causal horizons. Holographic screens exist generically, including in our universe, and regardless of the presence of black hole or de Sitter horizons.

Our proof, like Hawking's for his theorem, relied on the Null Energy Condition, that the stress tensor satisfies $T_{ab} k^a k^b\geq 0$ for any null vector $k^a$. As noted earlier, this condition can be violated in quantum field theory. Thus, it is natural to propose that the quantum-corrected area would satisfy a more robust monotonicity property. We called this conjecture the generalized second law for cosmology~\cite{BouEng15c}:
\begin{equation}
d S_\text{gen} >0. 
\end{equation}
(The strict inequality arises from a genericity assumption that simplifies the analysis.)

This is the first rigorous and covariant statement of a Second Law in cosmology. It does not, for example, rely on symmetries like homogeneity to subdivide a portion of the universe and consider its comoving entropy. Such a definition would at best be approximate and would not apply in general spacetimes.

Yet, our Second Law remains somewhat mysterious, in that we know of no thermodynamic argument for why it should hold. This is in marked contrast to Bekenstein's GSL, which was arrived at not by quantum-correcting Hawking's area theorem, but by demanding that the laws of thermodynamics hold in the exterior of black holes. (See Sections~\ref{sec-bhe}, \ref{sec-gsl} above.) It would be nice to fill this gap.

\subsection{Quantum Focussing Conjecture}

A consequence of Einstein's equation, Raychaudhuri's equation
\begin{equation}
  \theta' = -\frac{1}{2}\theta^2 - \varsigma_{ab}\varsigma^{ab} - 8\pi G\, T_{ab}k^a k^b
  \label{eq-raych}
\end{equation}
describes the evolution of a family (congruence) of null geodesics that form a null hypersurface~\cite{HawEll,Wald}. Here $\theta$ is the expansion, i.e., the logarithmic derivative of a small area element transported along the congruence. $\theta$ is defined in terms of the trace of the null extrinsic curvature, whose symmetric part gives the shear, $\varsigma_{ab}$. (We consider surface-orthogonal null congruences so there is no antisymmetric part or ``twist.'') The prime refers to a derivative with respect to the affine parameter along the geodesics. 

Assuming the null energy condition, $T_{ab}k^a k^b\geq 0$, Eq.~(\ref{eq-raych}) implies
\begin{equation}
  \theta'\leq 0~.
  \label{eq-classicalfocus}
\end{equation}
This is the classical focussing property of light: matter focusses but never anti-focusses light-rays. For example, initially parallel light-rays grazing the Sun are bent inwards, not outwards by the Sun's gravity.

If the null energy condition is violated, for example by Casimir energy or near a black hole horizon, then light-rays can antifocus, i.e., one can have $\theta'>0$. One can ask, however, whether a quantum-corrected version of Eq.~(\ref{eq-classicalfocus}) remains valid.

To find an appropriate formulation~\cite{BouFis15a}, we appealed to Bekenstein's generalized entropy, interpreting it as a quantum-corrected area $A_Q = 4G \hbar S_\text{gen}$. For any two-dimensional surface $\Sigma$ that divides the world into two sides, we defined the ``quantum expansion'' $\Theta$ analogous to $\theta$, as the response of the generalized entropy of $\Sigma$ to local deformations of the $\Sigma$ along an orthogonal light-ray.\footnote{Strictly, $\Theta$ is a functional derivative of the generalized entropy and so depends on $\Sigma$ globally. It reduces to the ordinary derivative in Eq.~(\ref{eq-classicalfocus}) as $G\hbar\to 0$, which has no such dependence.} It was then natural to conjecture that
\begin{equation}
  \Theta'\leq 0~.
  \label{eq-qfc}
\end{equation}
This is the {\em Quantum Focussing Conjecture} (QFC). See~\cite{BouFis15a} for precise definitions.

Some evidence for the validity of this conjecture comes from the fact that it reproduces other well-tested or proven statements as limiting cases. For example, if $\Sigma$ has $\Theta\leq 0$ initially, then Eq.~(\ref{eq-qfc}) implies the generalized covariant entropy bound~\cite{CEB1,FMW}, Eq.~(\ref{eq-gceb}). 
In particular, the QFC thus implies the Bekenstein bound~\cite{Bek81} in the strong form of Eq.~(\ref{eq-tight})~\cite{Bou02}.

The formulation of the QFC in terms of $S_\text{gen}$ provides a more rigorous statement of both bounds: the generalized entropy of the initial surface is greater than that of the final surface. (There exists another quantum version of the covariant bound which is defined only in the weakly gravitating limit~\cite{BouCas14a,BouCas14b}. In this regime, both versions can be applied, and there appears to be no logical relation between them. It would be nice to understand this better. Perhaps there exists a stronger statement that implies both?)

A further consequence of the QFC will be discussed in the next subsection.

\subsection{Quantum Null Energy Condition}
\label{sec-qnec}

The QFC led to a new result in quantum field theory, the {\em Quantum Null Energy Condition}~\cite{BouFis15a,BouFis15b}. The QNEC states that the local energy density of matter is bounded from below by an information-theoretic quantity:
\begin{equation}
  T_{ab} k^a k^b \geq \frac{\hbar}{2\pi} S_\text{out}'' ~.
  \label{eq-qnec}
\end{equation}
Here $T_{ab}$ is the stress tensor at a point $p$. We have implicitly chosen a surface $\Sigma$ that contains $p$ and divides the world into two parts, such that the classical expansion and shear at $\Sigma$ vanish.\footnote{An example of this is any infinite spatial plane in Minkowski space containing $p$. One can also consider more general settings~\cite{BouFis15b,Lei17,AkeCha17}.} The null vector $k^a$ is orthogonal to $\Sigma$ at $p$, and $S_\text{out}$ is the von Neumann entropy of the quantum fields on one side of $\Sigma$. The double-prime refers to a second functional derivative (with a Dirac delta-function stripped off), with respect to deformations of $\Sigma$ in the direction of $k^a$ at $p$.

The QNEC, Eq.~(\ref{eq-qnec}), follows immediately from the QFC by combining Eqs.~(\ref{eq-sg}), (\ref{eq-raych}), and~(\ref{eq-qfc}), and taking the limit as $G\to 0$. By construction, the initial classical expansion and shear vanish at $\Sigma$. By Eq.~(\ref{eq-raych}), the expansion will be $O(G)$ and proportional to the stress tensor on the null hypersurface orthogonal to $\Sigma$. Thus the classical piece, in this special setting, is of the same order as the quantum correction arising from the entropy. After cancelling out $G$, this results in Eq.~(\ref{eq-qnec}).

The QNEC is the first lower bound on the local stress tensor in quantum field theory. It implies the Averaged Null Energy Condition~\cite{WalYur91} on the integrated stress tensor, so the QNEC is stronger than the ANEC. Operationally, the QNEC can be thought of in terms of an observer who examines a signal or region of space up to a boundary. We can ask how the information gained by the observer would increase if they deformed the boundary to include a greater region. The QNEC tells us that the energy density of the signal limits the second derivative of the information content under this deformation.

We can integrate the QNEC twice, to obtain a precise version of Bekenstein's bound~\cite{Bek81}. Thus the QNEC is a stronger, differential version of the bound. It captures in the most local way possible the relation between energy, spacetime, and information that underlies both the Bekenstein bound and the holographic entropy bounds.\footnote{For interacting theories there is a precise sense in which this relation is actually an equality, $ T_{ab} k^a k^b = (\hbar/2\pi) S_\text{out}''$, namely when $\Sigma$ is deformed twice at the same point~\cite{LeiLev18}. The inequality in Eq.~(\ref{eq-qnec}) thus arises entirely from the off-diagonal contributions to the second functional derivative, when $\Sigma$ is defomed at two different points.}

The QNEC is purely a statement about quantum field theory, and it can be proven in that setting. This was done for free theories in Ref.~\cite{BouFis15b}, for interacting theories with a holographic dual in Ref.~\cite{KoeLei15}, and for general interacting theories in Ref.~\cite{BalFau17}. The field theory proofs are rather elaborate, totaling around 100 pages. By contrast, as noted earlier, it is trivial to obtain the QNEC from a conjecture about semiclassical gravity, the QFC.

This takes us back to the controversy surrounding the Bekenstein bound (Sec.~\ref{sec-controversy}). Unruh and Wald~\cite{UnrWal83} were right when they argued that quantum field theory already has properties sufficient to protect the GSL; in this sense, the Bekenstein bound was not ``needed.'' But the properties of QFT that guarantee this are somewhat obscure and conspiratorial from the QFT viewpoint. They can be much more directly understood as being required by the compatibility of field theory with gravity. This contrast is far more apparent today, as quantum gravity considerations have yielded more precise, provable new results in QFT, such as Eq.~(\ref{eq-qnec}) and~\cite{Cas08,Wal11,BouCas14a,BouCas14b}.

This exposes a major new facet of the unity of Nature, one that Bekenstein first glimpsed. Quantum gravity not only governs extreme regimes such as the beginning of the universe or the singularity inside a black hole. It also governs the information content of relativistic quantum fields at low energies. It will be important to probe this regime directly, and we will not have to wait for a Planck collider to do it.

\noindent {\bf Acknowledgments}~~~ This is a contribution to the {\em Jacob Bekenstein Memorial Volume}, to be published by World Scientific. I thank Slava Mukhanov and Eliezer Rabinovici for their encouragement and their patience. I am grateful to Stefan Leichenauer, Bob Wald and Aron Wall for detailed comments. My research is supported in part by the Berkeley Center for Theoretical Physics, by the National Science Foundation (award number PHY-1521446), and by the U.S.\ Department of Energy under contract DE-AC02-05CH11231 and award {DE-SC0019380}.

\bibliography{all}
\end{document}